\documentclass[10pt]{article}
\usepackage{amssymb}
\usepackage{amstext}

\newcommand*\de{\mathrm{d}}
 
\renewcommand*\epsilon{\varepsilon}
\renewcommand*\phi{\varphi}
\renewcommand*\theta{\vartheta}

\renewcommand*\AA{{}^{\text{\tiny T}}\!\! A}
\renewcommand*\P{{}^{\text{\tiny T}}\!\pi}
\newcommand*\PP{{}^{\text{\tiny TT}}\!\pi}
\newcommand*\HH{{}^{\text{\tiny TT} }\!h}
\renewcommand*\H{{}^{\text{\tiny T}}\!h}
\newcommand*\GG{{}^{\text{\tiny TT} }\!g}
\begin{document}
\title{\bf \large Hamiltonian reduction of spin-two theory and
of solvable cosmologies}

\author{ M. Leclerc\\ \small
  Section of Astrophysics and Astronomy,
Department of Physics,\\ \small  University of Athens, Greece}
 \date{\small February 10, 2007}
\maketitle

\begin{abstract}
The Hamiltonian reduction of the massless spin-two field 
theory is carried out following  the Faddeev-Jackiw approach. 
The reduced Hamiltonian 
contains only the traceless-transverse fields, but  not all 
of the non-propagating components can be determined by the constraints 
of the theory. The reason for this is found 
in the fact that the Hamiltonian is not gauge invariant. Consequences 
and implications for General Relativity are discussed and illustrated 
on the example of Robertson-Walker cosmologies with a scalar field. 
Also, it is shown that for those explicitely solvable models, the 
reduced form of the dynamics uniquely determines the operator ordering 
that has to be adopted in the Wheeler-DeWitt equation in order to maintain 
consistency. 
\end{abstract}

\section{Introduction}

It is generally believed that the free, massless spin-two field is 
entirely determined, in the Hamiltonian formulation, by the 
dynamical equations for the propagating (radiative) modes and by 
the constraints of the theory. Although this is supported by 
the naive counting argument (two propagating modes, four constraints 
and four undetermined gauge variables), we argue in this article   
that this assumption is actually mistaken. The reason can be 
recognized in different ways, for instance in the fact that 
the Hamiltonian is not gauge invariant off the constraint surface.  
A similar situation holds in General Relativity. Although the analysis is 
difficult to be carried out in the full theory, it is not hard to illustrate 
the corresponding issues by means of simple examples for which the Hamiltonian 
reduction can be carried out explicitely. A consequence of this particular 
feature is the fact that there exist spacetimes that are not classical, 
but are nevertheless free of dynamical quantum gravitational fields, 
 i.e., of gravitons. As a byproduct of our analysis, we show that 
the explicitely reduced theory dictates the operator ordering that has to 
be adopted  in the Wheeler-DeWitt equation in order to find results that are 
consistent with those one gets from the reduced theory. 

The article is organized as follows. In the next section, we deal with the 
special relativistic spin-two theory. In section \ref{gr}, arguments 
are presented that the situation can be expected to be similar in 
the framework of General Relativity. In the remaining  sections,   this is 
explicitely demonstrated on the example of Robertson-Walker cosmologies with 
 scalar field (section \ref{rob}), where we carry out the Hamiltonian 
reduction (section \ref{reduc}) and compare with the Wheeler-DeWitt 
approach (section \ref{wdw}), as well as with the results obtained 
from  fixation of the time coordinate  (section \ref{gf}). Finally, in 
section \ref{var}, we perform a change of variables such that the resulting 
theory is trivially reducible and the corresponding Wheeler-DeWitt 
equation is free of operator ordering problems.

\section{Spin-two theory}\label{spintwo}

We start from the first order Lagrangian (see,  e.g., \cite{leclerc} for 
its derivation  from the second order Fierz-Pauli Lagrangian)
\begin{equation}\label{1c}
 L = \pi^{\mu\nu} \dot h_{\mu\nu} -  H( \pi_{\mu\nu}, h_{\mu\nu}) 
+ h_{0 \mu} ( 2 \pi^{\mu\nu}_{\ \ ,\nu}) + 
h_{00}(-\frac{1}{2} h^{\mu\nu}_{\  \ ,\mu,\nu} 
+ \frac{1}{2} h^{,\mu}_{\ ,\mu}),   
\end{equation}
where our signature  convention is $\eta_{\mu\nu} =- \delta_{\mu\nu}$  
($\mu,\nu\dots = 1,2,3$), $h = h^{\mu}_{\ \mu}$. We will further use the 
notations $\Delta A = \partial^{\mu} \partial_{\mu} A= 
A^{,\mu}_{\ ,\mu}$ and $\Box A = \ddot A +  \Delta A  $, where a dot 
denotes the  time derivative. The  Hamiltonian has the form 
\begin{equation} \label{2c}
 H =  \pi^{\mu\nu} \pi_{\mu\nu} - \frac{1}{2} \pi^2 
+ \frac{1}{4} 
h_{,\mu} h^{,\mu} 
- \frac{1}{4} h^{\nu\lambda}_{\ \ ,\mu} 
h_{\nu\lambda}^{\ \ ,\mu}+ \frac{1}{2} 
h^{\mu\lambda}_{\ \ ,\mu} h_{\nu\lambda}^{\ \ ,\nu} 
- \frac{1}{2} h^{\mu\nu}_{\ \ ,\nu}h_{,\mu}. 
\end{equation}
In the first three sections of this article, 
integration over space is always understood 
and surface terms will be 
discarded whenever necessary. Note that we call {\it Hamiltonian} 
only the above expression, without the constraints coupled to it.
The absence of $h_{0\mu}$ and $h_{00}$ in the kinetic term of $L$ 
leads to the constraints 
\begin{equation}\label{3c}
2 \pi^{\mu\nu}_{\ \ ,\mu} = 0, \ \ - \frac{1}{2} 
h^{\mu\nu}_{\ \ ,\mu\nu} + \frac{1}{2} \Delta h = 0. 
\end{equation}
The Lagrangian (\ref{1c}) is 
easily shown to be invariant under the following gauge transformations 
\begin{eqnarray}\label{4c}
\delta h_{\mu\nu} &=& \xi_{\mu,\nu} + \xi_{\nu,\mu}, \\ \label{5c}
\delta \pi_{\mu\nu} &=& - \epsilon_{,\mu,\nu} + \eta_{\mu\nu} \Delta \epsilon,
\\ \label{6c}
\delta h_{0\mu} = \dot \xi_{\mu} + \epsilon_{,\mu},  && 
\delta h_{00} = 2 \dot \epsilon. 
\end{eqnarray}
This is a consequence of the invariance of the second order Lagrangian under 
$\delta h_{ik} = \xi_{i,k} + \xi_{k,i}$, ($i,k= 0,1,2,3$), with $\xi_0 = 
\epsilon$. It is needless to say that (\ref{4c}) and (\ref{5c}) are induced 
by the constraints (\ref{3c}). 
The reduction to the dynamical degrees of freedom consists 
in solving the constraints 
explicitely and bringing the Hamiltonian into a form 
that contains only the physical fields. This has been done in \cite{deser} 
starting from a slightly different (but equivalent) Lagrangian, and 
also in \cite{leclerc}, where we have reduced the spin-two Hamiltonian 
following along the lines of Faddeev and Jackiw \cite{faddeev}. The essential 
step is to express the fields $h_{\mu\nu}$ and the momenta $\pi_{\mu\nu}$ 
in terms of the traceless-transverse (TT) parts $\HH_{\mu\nu}$ and 
$\PP_{\mu\nu}$, and to recognize that, on the constraint surface, the 
remaining fields (longitudinal, trace) cancel out both in the kinetic 
term as well as in the Hamiltonian. In other words, on the 
constraint surface, we have  
\begin{equation}\label{7c}
L_{\text{\tiny R}} =  \PP_{\mu\nu} {}^{\text{\tiny TT}}\! \dot h^{\mu\nu} 
- H_{\text{\tiny R}}(\PP_{\mu\nu}, \HH_{\mu\nu}), 
\end{equation}
with 
\begin{equation}\label{8c}
H_{\text{\tiny R}} =  \PP^{\mu\nu}\, 
\PP_{\mu\nu} - \frac{1}{4} \HH^{\mu\nu}_{\ \ ,\lambda}
\HH_{\mu\nu}^{\ \ ,\lambda}. 
\end{equation}
The index is attached to $H_{\text{\tiny R}}$ (and $L_{\text{\tiny R}}$) 
to remind that this form 
of $H$ (and $L$) holds only if the constraints are satisfied. 
The equal time Poisson brackets can be read off from (\ref{7c})  and 
read $[\PP^{\mu\nu}(\vec x), \HH_{\alpha\beta}(\vec y)] 
= - \frac{1}{2}(\delta^\mu_\alpha 
\delta^{\nu}_{\beta} + \delta^{\nu}_{\alpha} \delta^{\mu}_{\beta}) 
\ \delta(\vec x - \vec y)$, and the dynamical equations 
$[H, \PP_{\mu\nu}] = -  {}^{\text{\tiny TT}}\! 
\dot \pi_{\mu\nu}$ and $[H, \HH_{\mu\nu} ] = - 
{}^{\text{\tiny TT}}\! \dot h^{\mu\nu}$ lead to 
\begin{equation} \label{9c} 
 - {}^{\text{ \tiny TT}}\! \dot \pi^{\mu\nu} = 
 \frac{1}{2}\Delta  \HH^{\mu\nu}, \quad
 -  {}^{\text{\tiny TT}}\! \dot h^{\mu\nu} =- 2\, \PP^{\mu\nu}. 
\end{equation}
which leads to the expected wave equation of the spin-two particle, 
 $\Box\, \HH^{\mu\nu}= 0$. 
As expected, the reduction leads straightforwardly to the
identification of the propagating field modes, and the transition to 
the quantum theory is now possible. However, it is obvious that 
the dynamical equations (\ref{9c}), together with the constraints 
(\ref{3c}) do not determine the system completely. In other words, 
(\ref{9c}) and (\ref{3c}) are not equivalent to the set of field 
equations that can be derived from the Lagrangian (\ref{1c}) 
(or, alternatively, directly from the second order Fierz-Pauli Lagrangian, 
see \cite{leclerc}). In fact, variation of $L$ with respect to 
$h_{\mu\nu}$ and $\pi_{\mu\nu}$ leads to field equations whose 
traces are respectively 
\begin{equation} \label{9cc}
\dot h + \pi - 2  h^{0\mu}_{\ \ ,\mu} = 0, \qquad 
- \dot \pi + \frac{1}{2} \Delta h - \frac{1}{2} h^{\mu\nu}_{\ \ ,\mu,\nu} 
+ \Delta h_{00}= 0, 
\end{equation}
 leading  to the configuration-space equation
\begin{equation} \label{10c} 
\ddot h + \Delta h_{00} - 2 \dot h^{0\mu}_{\ \ ,\mu} = 0, 
\end{equation}
 where we have used one of the  constraints already. 
Those  equations, which are gauge invariant,  cannot be obtained 
merely from the constraints (\ref{3c}) and the dynamical equations 
(\ref{9c}), as is easy to recognize (consider,  e.g., the configuration 
$\pi_{\mu\nu} = h_{\mu\nu} = h_{0\mu} = 0$, $h_{00} = \exp{(-r)}$, which 
solves (\ref{3c}) and (\ref{9c}), but is not  solution to (\ref{9cc}) or 
 (\ref{10c}). It is 
not hard to argue that there are no other  
 independent equations one 
obtains from (\ref{1c}) that are not contained in the constraints and in the 
dynamical equations, see \cite{leclerc}.  

We conclude that the spin-two field is not entirely characterized by the 
dynamical and constraint equations alone, in sheer contrast to 
conventional gauge theories, 
and we will now analyze both 
the reasons and the consequences of this observation. It is known that 
the Hamiltonian reduction sometimes leads to a loss of information. For 
instance, it has been demonstrated in \cite{pons} that this can happen in 
systems with ineffective constraints. Somewhat more close to our case 
are systems with so-called reducible constraints,  i.e., constraints 
that are not independent, see  \cite{barcelos} and \cite{banerjee}. 
The classical example is that of constraints of the form $p^{\mu\nu}_{\ \
  ,\mu} = 0$, with antisymmetric $p^{\mu\nu}$. Those constraints are 
interrelated via $p^{\mu\nu}_{\ \ ,\mu,\nu} = 0$ and therefore cannot 
generate independent gauge transformations. Here, we have a slightly 
different situation, namely the constraints 
$\frac{1}{2} G^{00} = \frac{1}{2}(\Delta h 
- h^{\mu\nu}_{\ \ ,\mu,\nu})$ and $G^{0\mu} = 2 \pi^{\mu\nu}_{\ \ ,\nu}$ 
satisfy the relation $\dot G^{00} + G^{0\mu}_{\ \ ,\mu} = 0$. This is 
a result of the Bianchi identities of the linear spin-two theory.  
(Recall that  $G^{00}=0$ and $G^{0\mu}=0$ are nothing but the field 
equations corresponding to $h_{00}$ and $h_{0\mu}$,  i.e., they are 
equal to the corresponding 
 components of the linearized Einstein tensor, 
expressed in phase-space variables. Note, however, the factor two, 
that results because $G^{ik}$ is defined upon variation with respect to 
$h_{ik}$, whose components are not the same than those obtained upon 
direct variation with respect to 
$h_{\mu\nu}, h_{0\mu}$ and $ h_{00}$, since 
in products like $a_{ik}a^{ik}$, 
the mixed components $a_{0\mu}$ occur twice.)
However, in contrast to the case of reducible constraints, 
the interrelation of the constraints manifests itself only on-shell, 
since we have to use  the explicit expressions for $\pi_{\mu\nu}$ in 
terms of $\dot h_{\mu\nu}$. 
Nevertheless, we retain the fact that,  strictly speaking, the constraints 
are not independent. 

A more obvious reason why the reduction must lead to the loss of 
a field  equation can be seen in the following. As is easily shown, 
the solutions to the second constraint in (\ref{3c}) can be written in 
the form $h_{\mu\nu} = \HH_{\mu\nu} + f_{\mu,\nu} + f_{\nu,\mu}$ (see 
\cite{deser} or \cite{leclerc} for the explicit expression of $f$), 
meaning that $h_{\mu\nu}$ is TT, up to a gauge
transformation. If we now choose the particular gauge $h_{\mu\nu} = 
\HH_{\mu\nu}$, and impose further $\pi  = 0$, which 
can be achieved with (\ref{5c}) with a suitable choice of $\epsilon$, 
then the first of  equations (\ref{9cc}) reduces to 
$h^{0\mu}_{\ \ ,\mu}=0$ and the second to $\Delta h_{00} = 0$. But those 
equations are now  in  the form constraints,  
i.e., they  do not depend on velocities. 
Thus, depending on the gauge one adopts, those   equations  appear 
either in the form of  constraints, either in the form of  dynamical 
equations.  Therefore, they  cannot be obtained as  dynamical equations 
of the form $[H, A] = - \dot A$, for some physical 
(gauge invariant) variable $A$, but nor can they  arise as  true 
constraints, at least not as long as we do not explicitely fix the gauge. 
Obviously, the Hamiltonian reduction process, which is based on solving 
the constraints and introducing 
Poisson brackets for the physical variables only, has no place for such an 
{\it intermediate} case. 

The mathematical reason for the failure of the reduction process 
is simply the fact that the Hamiltonian (\ref{2c}) is not gauge invariant 
off the constraint surface, in contrast to conventional 
gauge theories. Let us illustrate this on  the example of electrodynamics, 
where we have 
\begin{equation}\label{m1}
L = \pi^{\mu} \dot A_{\mu} - H  + A_0 \pi^{\mu}_{ \ ,\mu}, 
\end{equation}
with 
\begin{equation} \label{m2}
H = -\frac{1}{2} \pi_{\mu}\pi^{\mu} + \frac{1}{4}F_{\mu\nu} F^{\mu\nu}, 
\end{equation}
and we find 
$\delta (\pi^{\mu}\dot A_{\mu} + A_0 \pi^{\mu}_{\ ,\mu}) = 0$, as well 
as $\delta H = 0$ (under $\delta \pi_{\mu} = 0 $, $\delta A_{\mu} = 
\epsilon_{,\mu}$, $\delta A_0 = \dot \epsilon$). This means that 
the gauge dependent parts of $A_{\mu}$ (the longitudinal modes) are not 
contained at all in $H$, independently of whether we are on or off the 
constraint surface. The reduced Hamiltonian is given by 
\begin{equation} \label{m3}
H_{\text \tiny{R}} = - \frac{1}{2} \P_{\mu} \P^{\mu} + 
\frac{1}{2} \AA_{\mu,\nu} \AA^{\mu,\nu},   
\end{equation}
where $\AA_{\mu}$ is the transverse field. 
Note that for the magnetic field, we have  
$F_{\mu\nu} = A_{\nu,\mu} - A_{\mu,\nu} = \AA_{\nu,\mu} - \AA_{\mu,\nu}$, 
while  the electric field is given by $\pi^{\mu}$. Both are obviously 
gauge invariant, and moreover, they constitute the only gauge invariant 
quantities of the theory. (This can easily be generalized  to 
the case where  charged matter fields are present, see, e.g., \cite{faddeev}.) 
The point is that both the electric and the magnetic fields are completely 
determined by the dynamical variables $\P^{\mu}$ and $\AA_{\mu}$ and the 
constraint alone, since the only non-dynamical field occurring in 
those fields is the longitudinal component of $\pi^{\mu}$, which is 
fixed  by the constraint $\pi^{\mu}_{\ ,\mu} = 0$.

On the other hand, in the spin-two theory, we have a different situation. 
First, we note that  for  the 
transformations induced by $\xi_{\mu}$ according to  (\ref{4c})-(\ref{6c}) 
  we have again $\delta_\xi H = 0$ as well as  
$\delta_\xi (\pi^{\mu\nu} \dot h_{\mu\nu} 
+ h_{00} \frac{1}{2}G^{00} + h_{0\mu} G^{0\mu}) 
= 0$, where we use again the symbols $\frac{1}{2} G^{00}, G^{0\mu}$ for the 
expressions (\ref{3c}). This is just as in electrodynamics.  

Quite in contrast, however, for the transformations 
induced by $\epsilon$, although the total Lagrangian is invariant, 
we find 
\begin{equation} \label{m4}
\delta_\epsilon H = 2 \pi^{\mu\nu}_{\ \ ,\nu} \epsilon_{,\mu},
\end{equation} 
which shows that 
$H$ is not invariant off the constraint surface. Obviously, this variation is 
compensated by the variation of other terms in $L$.  Indeed, we find 
$\delta_\epsilon (H + h_{0\mu} 2 \pi^{\mu\nu}_{\ \ ,\nu}) = 0$.  

Thus, the 
Hamiltonian $H$, in contrast to its electromagnetic counterpart,  
is not independent of the gauge variables.   
As a result of this, although the total 
Lagrangian is (by construction) gauge invariant, we cannot simply 
discard the gauge variables (as can be done with the longitudinal 
parts of $A_{\mu}$), and simultaneously treat $h_{00}$ and $h_{0\mu}$ 
as arbitrary Lagrange multipliers  (as can be done with $A_0$). We 
can do either one or the other, but not both, contrary to what seems 
to be the believe in  \cite{deser}. 
On the level of the 
field equations, this is clearly demonstrated by the fact that those 
variables are interrelated by equations (\ref{9cc}) or (\ref{10c}).  
 If we discard the gauge variables and 
replace directly $h_{\mu\nu}$ and $\pi_{\mu\nu}$ by $\HH_{\mu\nu}$ and 
$\PP_{\mu\nu}$, then $h_{00}$ and (or) $h_{0\mu}$ cannot be arbitrary, but are 
determined by equation (\ref{10c}). Alternatively, if we fix the Lagrange 
multipliers to,  e.g., $h_{00} = h_{0\mu} = 0$, then (\ref{10c}),  
which is not identically satisfied on the constraint surface, determines 
the trace of $h_{\mu\nu}$.  We will see those features more explicitely in 
the framework of General Relativity later on. 

Since the variation of the Hamiltonian leads to a term proportional to 
the constraint, see (\ref{m4}), 
the above  described interrelation gets lost once we solve 
the constraints. 
This is the reason why the reduction process leads to a loss of information. 
It is important to remind once again that this information is of physical 
relevance and cannot be discarded, since the equations (\ref{9cc}) are gauge 
invariant. 

From the above considerations, one can conclude that there exists no 
Coulomb gauge for the spin-two field,  i.e., a gauge where all equations 
but the dynamical (TT) are free of time derivatives,  i.e., 
of the form of a Gauss type law. Indeed, as we have outlined above, 
it is possible,  e.g.,  to choose $h_{\mu\nu} = \HH_{\mu\nu}$ and 
$h^{0\mu}_{\ \ ,\mu} = 0$, such that field equations reduce to 
 $\Delta h_{00} = \Delta h_{0\mu} = 0$ (with solutions $h_{00} = 
h_{0\mu} = 0$) and $\Box \HH_{\mu\nu} = 0$. But this is not a Coulomb 
gauge. It is the analogous  of the radiation gauge $A_0 = 0$ and 
$A_{\mu}  = {}^{\text{\tiny T}}\!A_{\mu}$ in electrodynamics, which is,  
strictly speaking,  not a gauge, but rather  
a combination of the transverse gauge and the solution of Gauss' law 
$\Delta A_0 = 0$. The existence of this radiation gauge has been known 
at least since the works on gravitational waves by Einstein and Rosen, 
and is demonstrated in any textbook on General Relativity. For the 
Coulomb gauge, we need something more. The difference to the electromagnetic 
case lies in the fact that in Maxwell's theory, the transverse gauge 
can be imposed on $A_{\mu}$ independently of whether or not $A_{\mu}$ 
is a solution to the constraint (in fact, in phase-space, the 
constraint equation $F^{0\mu}_{\ \ ,\mu} = 0$ has the form $\pi^{\mu}_{\
  ,\mu} = 0$ and does not even depend on $A_{\mu}$). In the spin-two case, 
on the other hand, the TT-gauge is only possible 
if $h_{\mu\nu}$ is a solution of the constraint $h^{\mu\nu}_{\ \ ,\mu,\nu} 
- \Delta h = 0$. In the general case \cite{deser}, we have a decomposition 
$h_{\mu\nu} = \HH_{\mu\nu} + \H_{\mu\nu} + f_{\mu,\nu} + f_{\nu,\mu}$, 
where the transverse part $\H_{\mu\nu}$ cannot be gauged away. Only on 
the constraint surface, we have   $\H_{\mu\nu}= 0$. Why is this of 
importance? Well, suppose we couple the field to some matter field. Then 
the constraint changes, and so does the solution to the constraint. 
For instance, if we see in  the spin-two theory  the linearized 
approximation of General Relativity, then the constraint would take the 
form $G_{00} - \rho = 0$, where $\rho = T_{00}$ is a component of the 
stress-energy tensor of the matter fields. The solution to this 
constraint can be written in the form $h_{\mu\nu} = \HH_{\mu\nu} 
+ f_{\mu,\nu} + f_{\nu,\mu} + \eta_{\mu\nu}\,\rho/\Delta$, 
see \cite{leclerc}, 
and this is not gauge equivalent to $\HH_{\mu\nu}$ anymore. Thus, we cannot 
reduce $h_{\mu\nu}$ to its propagating components only. As a result, 
there is no Coulomb gauge, because even if we solve the four constraints 
(\ref{3c}), which can be viewed as the analogous to  Gauss' law, there will 
remain, apart from the dynamical equations, one more equation in the 
form of (\ref{10c}), eventually with sources, where $h$ cannot be eliminated. 

Summarizing, in contrast to the situation in Maxwell theory, 
the TT-gauge condition can only be imposed 
{\bf on} 
the constraint surface. In other words, the TT-gauge 
does not exist in the strict sense. There are only particular 
solutions (namely the radiative solutions) that can be brought into that 
form. For this reason, in textbooks on General Relativity, 
explicit reference 
to wave solutions has to be made in order to demonstrate the
TT nature of gravitational waves, quite in contrast to 
electrodynamics, where the transversality condition can be trivially 
 imposed right from the outset. 

A potential danger of those  specific features of the spin-two theories 
lies in the following. Suppose that,  for some reason, we work with a reduced 
number of degrees of freedom,  i.e.,  
we make some ansatz for $h_{\mu\nu}$,  
$\pi_{\mu\nu}$, $h_{0\mu}$ and $h_{00}$. Suppose further that 
 the number of independent functions we describe our fields with 
cannot be reduced any 
further by a gauge transformation. If we then solve the constraints, 
 we can  still not conclude that the remaining set of 
independent functions (that  are not 
fixed by the constraints) must correspond to 
 the physical variables of the theory, 
as would be the case in a conventional gauge theory. In particular, 
we cannot simply assume  that the unconstrained variables are 
subject to second quantization, because even when  the gauge is  fixed and 
the constraints are solved, there remains still one degree of freedom 
in the theory that is not a physical (propagating) field. 

If we assume for the moment that we have a similar situation in 
General Relativity (as we will see in the next sections), 
then the above remark turns out 
to be  of particular 
relevance. For instance, if we start from the outset with, e.g., 
a cosmological function and a scalar (matter) field only, then it is not 
at all obvious that those functions are indeed part of the dynamical  
fields and that, consequently, we should replace them with quantum 
operators satisfying canonical commutation relations. In fact, 
 the cosmological function (appearing as  conformal factor 
in front of the three-dimensional metric) is actually, in the weak field 
limit, related to the trace $h$ of the metric perturbation $g_{ik} = 
\eta_{ik} + h_{ik}$, which, however,  
appears in equation (\ref{10c}) rather than in the dynamical equations 
(\ref{9c}). Indeed, it turns out that the cosmological function is 
not a dynamical variable.. 

\section{General Relativity} \label{gr}

The important question is to what extend our considerations can be 
generalized to the self-interacting, generally covariant theory, 
to General Relativity. For the Hamiltonian formulation of General Relativity, 
we refer to \cite{dirac2,adm1,adm2,dewitt}. 
The explicit reduction of the theory is a highly difficult task, 
due to the non-linearity of the constraints. In the focus of the 
analysis is the identification of the physical variables. 
Their are again two dynamical degrees of freedom and we  will  
denote the corresponding pair of canonical variables with  
 $\PP^{\mu\nu}$ and $\GG_{\mu\nu}$, see \cite{adm1} and  \cite{adm2}. 
In fact, it has been argued  in \cite{adm2} that the similarity to 
the linear theory is actually quite strong, in particular with respect to 
the choice of the  canonical variables, since the non-linearity of the 
theory expresses itself mainly in the constraints, and not in the 
kinematical sector of the Lagrangian. The  first order Lagrangian is 
of the form \cite{dewitt}
\begin{equation}\label{11c}
L = \pi^{\mu\nu} \dot g_{\mu\nu} - N \mathcal H  - N_{\mu} 
\mathcal H^\mu, 
\end{equation}
where the constraints $\mathcal H = \mathcal H^{\mu}=0$ are equivalent to four 
of Einstein's equations, $G^{00}= G^{0\mu} = 0$, 
expressed in terms of $g_{\mu\nu}$ and 
$\pi^{\mu\nu}$ (see \cite{dewitt} for the explicit expressions). 
The Hamiltonian itself is zero (compare with (\ref{1c})), 
and thus, the equations for the physical fields are simply 
${}^{\text{\tiny TT}}\!\dot g_{\mu\nu}= {}^{\text{\tiny TT}}\!
\dot\pi^{\mu\nu} = 0$. Since we have not solved explicitely the constraints, 
we do not know exactly which parts of 
the fields correspond to the physical (TT) parts, but  we know that 
they are contained in $\pi^{\mu\nu}$ and $g_{\mu\nu}$ (and not in  
 $N,N_{\mu}$). Note that the non-TT parts of $\pi^{\mu\nu}$ and $g_{\mu\nu}$ 
do not necessarily have a vanishing time derivative, because, since 
those variables appear in the constraints,  for the derivation of 
the corresponding equations of motion, the full Hamiltonian 
$H  + N \mathcal H + N_{\mu}  \mathcal H^\mu$ (with $H = 0$) has to be used. 

If one  assumes that, as in conventional gauge theories, the 
configuration  is determined by the constraints
and by the dynamical equations alone, then,  since the latter are trivial, 
one comes to the conclusion that the complete information 
must be contained in the constraints alone. On the other hand, the linear 
theory shows that this is not necessarily the case. There might well 
be an additional equation, that is neither dynamical,  nor a constraint. 
Evidence that this is indeed the case comes from several directions. 
For one, there is of course the fact that there is a direct link 
between General Relativity and its linear counterpart. Perturbation theory 
should  to first order reproduce the results  that we find for the linear 
theory. Further, as a result of the Bianchi identity $G^{ik}_{\ \ ;k} = 0$, 
we see that the constraints are again on-shell related. 
One can also try to construct a metric such that the Einstein 
field equations contain second time derivatives of three independent 
functions, demonstrating in that way the fact that the corresponding set of 
first order equations contains at least three pairs of equations that are not
constraints, which is one more than there are physical degrees of freedom
 (see \cite{leclerc} for such an attempt). It turns out to be easier to 
demonstrate directly that the set of constraints and dynamical 
equations is underdetermined with  an explicit example. For instance, 
if we take $g_{\mu\nu} = \eta_{\mu\nu}$, 
$\pi^{\mu\nu} = N_{\mu} = 0$, and $N^2 = 
\exp(-r)$, then both the constraints $\mathcal H = 0$, 
$\mathcal H^{\mu} =0$ as well as the 
dynamical equations are trivially 
satisfied (the latter because we have $\dot g_{\mu\nu} = \dot \pi^{\mu\nu}=
0$, and thus, the same holds in particular for the TT parts, while for  the 
former, consider the explicit form of the constraints as 
given in \cite{dewitt}). Nevertheless, the above configuration is not a
solution to the complete set of the 
Einstein field equations in Hamiltonian form. 
This shows that there must 
be an extra equation apart from the constraints and the TT-equations. 
Indeed, one can directly verify that the equation  ${}^{4}\!R = 0$, with 
  ${}^{4}\!R $  the four dimensional curvature scalar, 
expressed in terms of $g_{\mu\nu}, \pi^{\mu\nu}, N$ and $N_{\mu}$, is 
not fulfilled for the above configuration. Although this equation 
contains gauge fields (non-TT parts), as well as $N$ and $N_{\mu}$, 
it cannot be argued that the equation is physically irrelevant. 
Those fields cannot be considered to be completely arbitrary, but 
rather transform in a well-defined way, such that altogether, 
the equation transforms into itself (after all, ${}^4 \!R$ is a scalar!).  

One can also argue as follows. If one  considers 
again the trace of the Einstein field 
equations, ${}^{4}\!R = 0$, 
(the explicit expression for ${}^{4}\!R$ can be found in \cite{dewitt}),  
then one recognizes that the only time derivative is  contained in 
a term of the form $(g_{\mu\nu} \pi^{\mu\nu})\dot{}$, or simply $\dot \pi$. 
Therefore, if we choose the gauge $N = 1, N_{\mu} = 0$, then the 
equation ${}^{4}\!R = 0$ 
contains velocities, namely $\dot \pi$. Choosing instead 
the gauge $\pi = 0$ (e.g., together with $(\sqrt{g}\ g^{\mu\nu})_{,\mu}= 0$), 
then the same equation does not contain velocities. Thus, we have again 
the situation that there are certain equations that can appear either in 
the form of a dynamical equation, or in the form of a constraint, 
depending on the gauge one adopts. This is similar to the case of 
equations (\ref{9cc}) or (\ref{10c}). 
As we have outlined above, such {\it intermediate} 
equations are lost during the  reduction process. We should point out, 
 however, that there are certain restrictions concerning the above 
gauge choices. For instance, it has been argued in \cite{dewitt} that 
 the gauge $\pi = 0$ can be imposed in asymptotically flat spacetimes, but not 
 in finite spaces with non-vanishing curvature. 
(It is actually quite a general feature of General Relativity that the 
procedure frequently depends right from the start on the solutions 
one wishes to obtain at the end of the day. We will encounter this 
feature in more detail in the next sections.) 

Summarizing, there seems to be strong evidence that the constraints 
 (together with the trivial dynamical equations) do not 
contain the complete physical information and that there is 
again some kind of interrelation between $N, N_{\mu}$ (or $g_{0\mu}, g_{00}$) 
and some (or one) of the gauge variables contained in the non-TT parts of 
$\pi^{\mu\nu}$ and $g_{\mu\nu}$. 

Nevertheless, one should not take the above considerations too serious. 
There are fundamental issues that we have not discussed. For instance, 
the relation to the linear theory is not really as straightforward   
as one might think. This is already obvious from the fact that 
the {\it linearized} Hamiltonian (\ref{2c}) can certainly not be obtained by 
linearization of the full Hamiltonian of General Relativity, which is zero, 
but rather emerges from the second order terms of the constraint $\mathcal H$, 
taking $N = 1+h_{00}/2$.  
Also, there is the matter of the 
occurrence of a surface term in (\ref{11c}) that 
we have omitted for simplicity. 
Another issue concerns the difference of the boundary conditions 
between  conventional gauge theories and General Relativity. For instance, 
one could argue that, in contrast to electrodynamics,  the latter admits 
solutions that are fundamentally non-static and nevertheless are free of 
radiation (cosmology).  Most importantly, however, 
there is a fundamental difference between the coordinate transformations 
of General Relativity and the corresponding gauge transformations 
in the linear theory. In that context, there are at least two issues that 
have to be considered, namely the  construction of 
physical (gauge invariant) quantities and the special role of the time 
coordinate. 
Those issues have been intensively studied  in 
the literature on quantum gravity, and it remains to see to what extend 
 they affect our specific discussion. 
  
In any case, it turns out that for simple models, like the 
Robertson-Walker cosmologies  we will 
analyze in the next sections, the  situation is the same  
as in the spin-two theory, meaning that there is again an equation 
that is neither part of the dynamical equations, nor of the 
constraints, and is nevertheless physical in the sense that it is 
necessary in order to describe the configuration completely.

\section{Robertson-Walker cosmologies} \label{rob}

We start from the Lagrangian
\begin{equation}\label{1}
L = - \frac{6R}{N} \dot R^2 + \frac{R^3}{2N} \dot \phi^2, 
\end{equation}
which leads to the field equations for a homogeneous,  massless  
scalar field $\phi = \phi(t)$ in the spatially flat 
Friedman-Robertson-Walker spacetimes 
\begin{equation} \label{2}
\de s^2 = N(t)^2 \de t^2 - R(t)^2 \delta_{\mu\nu} \de x^{\mu} \de x^{\nu}.  
\end{equation}
For details, we refer to the detailed study carried out in \cite{isham}. 
For mathematical simplicity, we confine ourselves to flat spaces $k=0$ 
and to the massless case $m=0$, which are explicitely solvable. 
Obviously, the system of field equations obtained from (\ref{1}) is 
underdetermined and has to be supplemented by a gauge choice, e.g., 
$N=1$. 

The first order form of the Lagrangian reads 
\begin{equation}\label{3}
L = \pi_R \dot R + \pi_{\phi} \dot \phi + N 
(\frac{\pi_R^2}{24 R} -\frac{\pi_{\phi}^2}{2 R^3}). 
\end{equation}
For the complete analysis of the solutions,  we refer 
to \cite{isham}. However, since we will repeatedly have to take square 
roots of certain variables, it is useful to have in mind the signs 
of the canonical variables and of the velocities. First of all, 
we can assume $N>0$ and $R>0$ (since the original 
variables, entering the metric, were  $N^2$ and $R^2$ anyway). 
This leaves us essentially with four cases to be considered, 
which, for our purpose, are  conveniently characterized by the signs of 
$\pi_{\phi}$ and $\pi_R$,  
\begin{eqnarray} \label{4}
\pi_R >0, \pi_\phi >0: \qquad \dot R < 0, \dot \phi > 0 \\
\pi_R >0, \pi_\phi <0: \qquad \dot R < 0, \dot \phi < 0 \label{5}\\
\pi_R <0, \pi_\phi >0: \qquad \dot R > 0, \dot \phi > 0 \label{6} \\
\pi_R <0, \pi_\phi <0: \qquad \dot R > 0, \dot \phi < 0. \label{7}
\end{eqnarray}
We will mostly deal with case (\ref{6}), but from a mathematical point of 
view, the four cases are quite similar. 

We note that the action 
$S= \int L \de t $ is invariant under time reparameterizations 
$t \rightarrow \tilde t$ if $N$ transforms accordingly, i.e., such that 
$N \de t $ is invariant. What is more interesting for our analysis is 
the invariance under the gauge transformation induced by the constraint 
that results from variation of $L$ with respect to $N$, namely 
\begin{equation} \label{8} 
G = \frac{\pi_R^2}{24 R} -\frac{\pi_{\phi}^2}{2 R^3} = 0. 
\end{equation}
Introducing canonical Poisson brackets $[\pi_R,R] = [\pi_{\phi},\phi] = -1$, 
we find that (\ref{8}) generates the following transformations on the 
fields and momenta
\begin{equation}\label{9}
 \delta R = - \frac{\pi_R}{12 R} \epsilon, \quad \delta \phi =
\frac{\pi_{\phi}}{R^3} \epsilon, \quad \delta \pi_R = (- \frac{\pi_R^2}{24 R^2}
+ \frac{3 \pi_{\phi}^2 }{2 R^4})\epsilon,\quad   \ \delta \pi_{\phi} = 0, 
\end{equation}
where $\epsilon$ is an infinitesimal parameter. Indeed, the Lagrangian 
(\ref{3}) is invariant under (\ref{9}) provided that 
 $N$ transforms according to 
$\delta N = \dot \epsilon$. The above transformation is the remnant  
of the coordinate transformations $\delta g_{ik} = g_{ik,m}
\xi^{m} + \xi^{m}_{\ ,k} g_{im} + 
+ \xi^{m}_{\ ,i} g_{mk} $ and $\delta \phi = \phi_{,m}
  \xi^{m}$ of the full theory.  

In the following section, we will perform the Hamiltonian reduction 
of the theory (\ref{3}) and compare the results, 
in section \ref{wdw}, with the Wheeler-DeWitt  approach as well as 
with the results obtained from an explicit choice of 
the time coordinate (section \ref{gf}) and by a change of variables (section 
\ref{var}).

\section{Explicit reduction} \label{reduc}

Just as in section \ref{spintwo}, we will identify the dynamical 
variables of the  theory by the reduction process 
 of Faddeev and Jackiw \cite{faddeev}, which relies 
on solving the constraint explicitely and obtain a first order Lagrangian 
with a reduced number of variables. In our case, it turns out to 
be convenient to solve (\ref{8}) with respect to $\pi_R$ (although 
other choices are possible). We find 
\begin{equation} \label{10}
\pi_R = \pm \frac{\sqrt{12}}{R} |\pi_{\phi}|, 
\end{equation}
which requires again the differentiation between the four cases 
(\ref{4})-(\ref{7}). The Lagrangian is easily put into the 
form 
\begin{equation}\label{11}
L = ( \phi \pm \sqrt{12} \ln R)\dot{} \  \pi_{\phi},
\end{equation}
where  the minus sign (in the following 
referred to as case 1) holds for  the cases (\ref{5}) and (\ref{6}) 
and the plus sign (case 2)  for the cases (\ref{4}) and (\ref{7}). 
(For simplicity, 
we will not notationally differentiate  between the reduced and unreduced 
Lagrangians in the following sections.)  

It is convenient to  introduce the variables 
\begin{equation} \label{12}
X= \phi - \sqrt{12} \ln R, \ \ Y = \phi + \sqrt{12} \ln R. 
\end{equation}
Let us confine ourselves to case 1. Then we have 
\begin{equation} \label{13} 
L = \dot X \pi_{\phi}, 
\end{equation}
that is, the dynamical variables are $X$ and $\pi_{\phi}$, and 
the corresponding Poisson brackets are given by 
\begin{equation}
[\pi_{\phi},X] = -1, \qquad [X,X]= [\pi_{\phi},\pi_{\phi}]=0. 
\end{equation}
The Hamiltonian is zero and the 
 classical equations of motion are obviously $\dot \pi_{\phi} = 0$ 
and $\dot X = 0$. We see that, as has already been stated 
by Dirac \cite{dirac2}, 
finding the dynamical variables and solving the field equations 
is essentially one and the same thing in covariant theories.

Since the constraint has been solved for $\pi_R$, we can now express 
the transformations (\ref{9}) in the form 
\begin{equation}\label{14}
\delta R = \frac{\pi_{\phi}}{\sqrt{12} R^2} \epsilon, \qquad 
\delta \phi = \frac{\pi_{\phi}}{R^3} \epsilon, \qquad \delta \pi_{\phi} = 0, 
\end{equation} 
and find 
\begin{equation} \label{15}
\delta X = 0, \qquad \delta  Y = 2 \frac{\pi_{\phi}}{R^3} \epsilon, 
\end{equation}
which confirms that $X$ and $\pi_{\phi}$ are indeed gauge invariant, as 
they should, while $Y$ is a gauge variable. 
(Note that in case 2, the role of $X$ and $Y$ are reversed.)

It is already obvious at this stage that the situation is similar 
to the spin-two case. Indeed, from  
the reduced dynamics that follow from the Lagrangian (\ref{13}), 
we can (trivially) determine $X$ and $\pi_{\phi}$
(just like  $\AA_{\mu}$ and $\P^{\mu}$ in electrodynamics) 
while  the constraint determines $\pi_R$ (just as it determines 
 the longitudinal part 
of $\pi^{\mu}$ in electrodynamics). Further, the variable $Y$ is a gauge 
variable and not of physical interest (just like the longitudinal part 
of $A_{\mu}$ in electrodynamics). However, from the reduced Lagrangian 
and from the constraint alone, we cannot determine $N(t)$. This 
is again similar to electrodynamics, where $A_0$ cannot be determined 
from the reduced Lagrangian. There is, however, a fundamental difference. 
In electrodynamics, we do not need $A_0$. As we have outlined previously, 
once we have determined both the 
longitudinal and the transverse parts of $\pi^{\mu}$ 
(i.e., of the electric field), 
the first one by the constraint and the second one by the dynamical 
field equations, and having in addition determined the transverse part of 
$A_{\mu}$  by the dynamical equations, leading to the magnetic field, 
we have already determined the complete set of gauge invariant quantities. 
There is nothing else we need to know. On the other hand, in the present 
 case, even if we fix the  variable $Y$ by choosing a gauge, 
we still need 
to know $N(t)$ in order to determine the spacetime geometry completely. 
Alternatively, we could fix $N$ to, say $N = 1$, leaving us with 
an undetermined $Y$, and again to an undetermined geometry (since 
we need both $X$ and $Y$ to determine $R$). 

In other words, there are physically relevant quantities that cannot 
be determined by the dynamical equation and the constraints alone. 
It is not hard to show that, for instance, the 
trace of the Einstein equations, $G^{i}_{\ k} = T^{i}_{\ k}$ 
involves both $Y$ and $N$. This equation, which can be derived from 
(\ref{3}), but not from the reduced Lagrangian, still remains non-trivial, 
 even if we fix the gauge by fixing $N$ or $Y$ (explicitely or implicitly). 
More explicitely, the equation obtained from variation of (\ref{3}) 
with respect to $R$, using $\dot \pi_{\phi} = 0$ and the 
expression for $\pi_R$ (here for case 1), can be brought into the form 
\begin{displaymath}
0 = \pi_{\phi} \sqrt{12} R^2 \dot R  - N \pi_{\phi}^2, 
\end{displaymath}
and is easily shown to be invariant under (\ref{14}) and 
$\delta N = \dot \epsilon$. Even if we 
fix $N$ to, e.g.,  $N=1$, the equation still contains both $X$ and $Y$, 
meaning that it is not determined by 
the dynamical variables $X$ and $\pi_{\phi}$ only. A similar situation holds 
for the gauge choice, e.g.,  $R = t$. 

Thus, just as in 
the linear spin-two theory, we have a physical (i.e., 
gauge invariant) relation between non-dynamical variables, namely between  
$N$ and $Y$. Another line of argumentation is to consider the 
equation $\dot Y = N \frac{2 \pi_{\phi}}{R^3}$, which can also be 
derived from (\ref{3}). Since $Y$ is a gauge variable, $\dot Y$  can be 
transformed to a constant (or at least to some 
expression of $t$ and of the field variables, see section \ref{var}). 
Thus, depending on 
the gauge one adopts, this equation can occur both in the form 
of a constraint (i.e., not involving velocities), or  in the 
form of a (seemingly) dynamical equation for $ Y$. As outlined 
previously, the reduction process, which is based on solving 
the (true) constraints and reducing the theory to the truly 
dynamical (gauge invariant) variables, has no place for such an intermediate 
case. Reference to the initial, unreduced Lagrangian can therefore not 
be avoided. We will see the consequences of this feature later on.

Let us return to the reduced Lagrangian (\ref{13}). Having identified 
the dynamical variables, the transition to the quantum theory is 
straightforward. We replace the  dynamical variables with 
operators satisfying the commutation relations 
\begin{equation}\label{16}
i[\pi_{\phi}, X] = 1,\qquad 
 i [X,X] = i[\pi_{\phi},\pi_{\phi}] = 0,
\end{equation}
which can be explicitely realized with $X$ a multiplication 
operator and 
\begin{equation} \label{17}
\pi_{\phi} = - i \frac{\partial }{\partial  X}. 
\end{equation}
Since the Hamiltonian is zero, the Schroedinger equation on the state 
functional (which, in the homogeneous case, is simply a wave function), 
$H \psi(X,t) = i \frac{\partial }{\partial t} \psi(X,t)$,  reduces 
to $\frac{\partial }{\partial t} \psi(X,t) = 0$, i.e., $\psi$ does not 
explicitely depend on the time coordinate. In other words, the wave 
function may be any function of $X= \phi - \sqrt{12} \ln R$, 
\begin{equation} \label{18}
\psi = \psi(X) = \psi(\phi-\sqrt{12} \ln R), 
\end{equation}
subject to suitable boundary conditions. That is the main result of 
this section. Obviously , in case 2,  
where $\pi_{\phi}$ and $\pi_R$ are both of the same sign, we 
have a similar result, namely $\psi = \psi(Y) = \psi(\phi + \sqrt{12} 
\ln R)$. 

Equation (\ref{17}) is all we get from the canonical procedure. Everything 
else has to be obtained from purely physical considerations. This concerns 
in particular the exact definition of the Hilbert space and the boundary 
conditions, but even if this is done, there remains the fundamental question 
which $\psi$ one actually should choose in order to get a reasonable
description of the situation. It can hardly be expected that $\psi$ is 
completely determined  by the boundary conditions alone. 

As a final remark, we note that there is only one quantum degree of 
freedom in the theory. This is of course the result of the very 
restricted ansatz (\ref{2}) and $\phi = \phi(t)$. Whether we attribute 
this degree of freedom to the gravitational field  or to the scalar field 
is a matter of convention (recall that $X = \phi - \sqrt{12} \ln R$), 
although the attribution to the scalar field seems more natural.
(One can argue, for instance, that in the absence of the scalar field, 
no dynamical degree of freedom would remain in the theory.) 
In the full theory (General Relativity with scalar field), 
we would end up with  three pairs of canonically conjugate dynamical 
variables and attribute two of them to the gravitational field. 
Convenient examples to study such systems explicitely are given by 
 the generalized homogeneous cosmologies described, e.g.,  
in \cite{misner}. 

There is one important point we have omitted in our discussion. Solving 
equation (\ref{10}), we have assumed that we deal either with one 
or with the other type of classical solutions. The corresponding 
quantum theory is described by a wave function either in the 
form $\psi(X)$ or in the form $\psi(Y)$. However, there is 
no need to restrict the  quantum theory to a specific type 
of classical solutions (although, of course, a very important restriction 
has already been made by starting with the specific metric (\ref{2})). 
This means that actually, we have to admit both type of solutions. In 
other words,  
invoking the superposition principle, we can conclude that the 
most general wave function is of the form $\psi = \psi_1(X) + \psi_2(Y)$. 
Therefore, strictly speaking, the reduction procedure is not completely 
consistent in the way it has been presented above. In particular, 
if we allow for those mixed type of solutions, we cannot actually 
identify the physical variables. For instance, the gauge transformations 
(\ref{9}) cannot be written in the form (\ref{14}) (because the 
solution of the constraint is not unique), and thus, 
the variation of $X$ and $Y$ under gauge  transformations will 
be proportional to the two roots of the constraint equation, meaning 
that it remains unspecified which of both variables is actually gauge 
invariant. As a result, one cannot reduce the theory to 
its two dynamical degrees of freedom $\pi_{\phi}$ and $X$ 
and should rather work with four operators instead. In this sense,  
the Wheeler-DeWitt approach, to which we turn now, is preferable, because 
it does not require us to determine the classical roots of  
the constraint equation.

\section{Operator ordering in the Wheeler-DeWitt equation} \label{wdw}

We now compare the  results of the previous section with the conventional 
approach to canonical gravity, where the constraints are 
not solved, but imposed  on the quantum states \cite{dirac2, dewitt}. 
We thus treat $R, \pi_R$ and $\phi, \pi_{\phi}$ as canonical pairs of 
variables and introduce the corresponding quantum operators, 
choosing again multiplication operators for $\phi$ and $R$, as well as 
$\pi_R = - i \frac{\partial}{\partial R}$, $\pi_{\phi} = - i \frac{\partial }
{\partial \phi}$. 
As before, the Hamiltonian (on the constraint surface) is zero, and thus, 
the wave function $\psi(R,\phi)$ does not  explicitely depend on 
the time coordinate. The dynamics are now described by the 
so-called Wheeler-DeWitt equation obtained 
by imposing the constraint (\ref{8}) on the state  
\begin{equation} \label{19}
\left(\frac{\pi_R^2}{24 R} 
-\frac{\pi_{\phi}^2}{2 R^3}\right)\ \psi(R,\phi) = 0. 
\end{equation}
A major problem concerns the operator ordering in the first term. 
This issue, in the general framework of canonical gravity, 
 has been addressed many times in literature, see 
\cite{anderson,komar,christo2,christo3,tsamis,christo}, and solutions 
have been proposed, based on hermiticity arguments and the requirement 
that the quantum algebra of the constraints be isomorphic to the 
classical algebra. To end up with a unique factor ordering, the 
authors of \cite{christo2,christo3} further require invariance under 
field redefinitions  $g_{\mu\nu} 
\rightarrow \tilde g_{\mu\nu}(g_{\lambda \kappa})$. A common 
feature of those argumentations is that the justification of a specific 
choice of factor ordering is based on the explicit construction of 
a scalar product in the  Hilbert space of the state functionals, 
since prior to this,  any discussion on hermiticity is meaningless.  

However, in our specific case, it turns out that, in view of the results 
of the previous section, the order of the operators occurring in (\ref{19}) 
is already determined uniquely, and no further arguments are required. 
Indeed, if the Wheeler-DeWitt equation is required to do what it is 
supposed to do, namely to eliminate the non-physical degrees of freedom 
from the theory  (and not to impose conditions on the physical degrees 
of freedom) then it should be identically satisfied for the solutions 
of the state function derived in the previous section, 
which are physical by construction.   

For simplicity, we restrict ourselves again to case 1, 
 where $X$ is 
the physical variable and $Y$ the gauge variable, but the 
analysis can be straightforwardly generalized to the mixed case. 
According to the results of the previous section, 
any function of $X$ should identically satisfy the Wheeler-DeWitt equation. 
Expressed in the variables $R,\phi$, this means that 
\begin{equation} \label{20}
\left(\frac{\pi_R^2}{24 R} -\frac{\pi_{\phi}^2}{2 R^3}\right)\ 
\psi(\phi - \sqrt{12}\ln R) 
\equiv 0 
\end{equation}
identically, with $\pi_R = - i \frac{\partial }{\partial  R}$. It is not hard 
to show that, up to equivalent orderings, this requires the Wheeler-DeWitt 
equation to be written in the form 
\begin{equation} \label{21}
\left(\frac{1}{24 R^3}(R  \pi_R R \pi_R)
- \frac{\pi_{\phi}^2}{2R^3}\right)
\ \psi(R,\phi) = 0.  
\end{equation}
An equivalent ordering is, for instance, 
$\left(\frac{1}{24}(2 \frac{1}{R} \pi_R^2 - \pi_R \frac{1}{R}\pi_R)
- \frac{\pi_{\phi}^2}{2R^3}\right)
\ \psi(R,\phi) 
= 0. $ Any other (i.e., inequivalent) ordering would lead to a quantum  
theory that is inequivalent to the theory obtained from the direct 
reduction of the Lagrangian 
to its physical degrees of freedom, as described in section \ref{reduc}. 
The ordering (\ref{21}) can  indeed be found in literature, 
see for instance \cite{page}, but on the other hand, 
it differs, e.g.,  from the choice adopted in \cite{dewitt}, 
which is $\sim R^{-1/4} \pi_R R^{-1/2} \pi_R R^{-1/4}$, as well as 
from the one adopted in \cite{christo4}, which is $\sim R^{-1/2} \pi_R 
R^{-1/2} \pi_R$. Not only is  (\ref{20})  not  identically fulfilled 
 with such orderings, but  in fact, it  does not 
admit any solutions of the form $\psi(\phi - \sqrt{12} \ln R)$ at all.  

Once we have fixed the operator ordering, we can explicitely 
solve the Wheeler-DeWitt equation, and we find the general solution
\begin{equation} \label{22}
\psi(R,\phi) = \psi_1 (R- \sqrt{12} \ln R) + \psi_2(R+ \sqrt{12} \ln R), 
\end{equation}
where $\psi_1$ and $\psi_2$ are arbitrary functions of their 
arguments. In other words, 
$\psi = \psi_1(X) + \psi_2(Y)$. As expected,  the 
Wheeler-DeWitt equation does not distinguish between case 1 (where $X$ is 
the dynamical variable) and case 2 (with $Y$ dynamical), which 
arose upon choosing between the roots (\ref{10}) of the constraint, 
since, just as the classical constraint, the Wheeler-DeWitt equation is 
insensitive to a sign change of $\pi_R$ or  $\pi_{\phi}$.  But, as 
we have pointed out at the end of the previous section, this 
is actually a good feature, because the reduction process is not 
unambiguous and therefore, one  should allow for both type of 
solutions $\psi(Y)$ and  $\psi(X)$, and thus also for superpositions 
$\psi = \psi_1(X) + \psi_2(Y)$. 

On the other hand, since an initial solution $\psi(X)$ will 
remain in this form at all times, we can also consider the 
restriction of the theory to case 1, as we did in the previous section. 
Then, we known that $X$ is the physical variable, and that 
the solutions $\psi(Y)$ have to be excluded. Since this cannot be 
done by the Wheeler-DeWitt equation, one will have to impose  
suitable boundary conditions on equation 
(\ref{21}) such that only the solutions $\psi= \psi(X)$ survive. 

Another way is to allow for all the solutions, but to take care that 
the {\it unwanted} solutions $\psi(Y)$ do not contribute to 
physical quantities (e.g., expectation values).  
Indeed, it turns out that,  
if physical equivalence to the approach of the previous section 
is to be obtained, the construction of the scalar product in the 
Wheeler-DeWitt approach is essentially 
fixed. In the reduced theory described by (\ref{13}), an obvious choice 
 consists in 
\begin{equation} \label{23}
<\psi_1(X), \psi_2(X)>\ =  \int \psi_1^* \psi_2 \de X, 
\end{equation}
where the integration can be taken over the complete set of reel numbers
 (or an appropriate subset, we do not deal with the details here, but 
confine ourselves to a brief outline of the  line of argumentation). 
Note that if we impose 
as boundary conditions that the wave functions $\psi$ vanish
at the boundary $X = \pm \infty$ (or any appropriate boundary), 
then the momentum $\pi_{\phi} = 
- i \frac{\partial}{\partial X}$ is hermitian. The scalar product 
(\ref{23}) is trivially invariant, since the gauge variables have 
already been eliminated from the theory. 

On the other hand, in the Wheeler-DeWitt approach, one could start with 
\begin{equation} \label{24}
<\psi_1(R,\phi),\psi_2(R,\phi)> 
\ = \int \psi_1^*(R,\phi) \psi_2(R,\phi) g(R,\phi) \de R \de \phi, 
\end{equation}
where the integration is carried out over a suitable set 
(e.g., positive reel numbers for $R$ and reel numbers for $\phi$), and 
$g(R,\phi)$ is a weight function that is chosen such that the scalar 
product is gauge invariant. Performing a change of variables, we can 
write   
\begin{equation}\label{25}
<\psi_1(X,Y),\psi_2(X,Y)> 
\ = \int \psi_1^*(X,Y) \psi_2(X,Y) \tilde g(X,Y) \de X \de Y, 
\end{equation}
For the particular functions $\psi = \psi(X)$,  
the scalar product (\ref{25})  reduces to the form  
 (\ref{23}) if we choose $\tilde g= \tilde g(Y)$ 
satisfying $\int \tilde g(Y) \de Y = 1$.  

The major problem with such  constructions is the fact that they are  
only valid for the theory that is confined to the classical solutions 
of type 1. In case 2, the role of $X$ and $Y$ are to be reversed. 
However, since there is no need for the physical wave solutions to be in 
a pure state $\psi(X)$ (or $\psi(Y)$), we have to generalize to the 
case where both $\psi(X)$ and $\psi(Y)$ can contain physical contributions. 
In other words, since we actually cannot specify whether the 
physical variable is $X$ or $Y$,  we cannot simply exclude 
one or the other set of solutions of the wave function. A scalar 
product for mixed functions $\psi(X,Y)$ 
that reduces to the above in the pure cases $\psi = \psi(X)$ or  
$\psi  = \psi(Y)$ can be constructed in the form  
\begin{equation}
 < \psi_1(X,Y), \psi_2(X,Y)>\ = \int \psi^*_1(X,Y) \psi_2(X,Y)
[\delta(X-x_0) + \delta(Y- y_0)] \de X \de Y, 
\end{equation}
where $x_0$ and $y_0$ are suitable constants (or eventually infinite) 
and  from the solutions of the Wheeler-DeWitt equation, we can select 
 the physical parts by imposing suitable boundary conditions 
(e.g., imposing $\psi(x_0, Y) = 0$ leaves us with $
<\psi_1 , \psi_2>\ = \int \psi_1^*(X,y_0) \psi_2(X,y_0) \de X$, which 
is of the form (\ref{23})). 

The relation to the invariant scalar products 
presented in literature, see, 
e.g., \cite{dewitt,christo,tsamis}, is not obvious and remains to 
be examined. In particular, we should note that it is not at all 
obvious that one should start with the form (\ref{23}) in the reduced 
theory. Other choices are possible and lead to different results 
in the corresponding Wheeler-DeWitt product, with, e.g.,  
$\rho(R,\phi)$ given in terms of a differential operator instead of 
a function. The choice of the scalar product is a highly non-trivial 
matter and we refer to the literature for details, see, e.g.,
\cite{vilenkin}. Our intention is merely to show that one could 
start the discussion from the explicitely reduced theory, whose structure 
is much simpler, and then try to derive the corresponding form of the 
scalar product in the Wheeler-DeWitt approach by requiring consistency 
with the reduced theory.  

Although in the full theory, the complete reduction cannot 
be performed explicitely, there is an alternative way to reduce the theory to 
its physical degrees of freedom, which is  by fixing the gauge. 
Again, this is 
not really a trivial matter, and in general, the allowed gauges can 
depend on the topological properties of the solutions one wishes to 
obtain, 
but at least one does not need the explicit solutions, in contrast to 
the reduction approach of section \ref{reduc}. Therefore, in the next 
section, we will illustrate this procedure for our simple model and 
compare the results with those obtained until now.

\section{Time coordinate fixation}\label{gf}

There are several natural choices for the time coordinate, all having  
 advantages and  limitations. We refer to \cite{isham} for a more detailed 
analysis and confine ourselves to a brief outline 
of two specific examples. In order to avoid case differentiations, 
we assume, throughout this section, that the solutions are of 
type (\ref{6}), that is, of case 1 with $\dot R >0$. 

First, we choose $R = t^{1/3}$. Before we reduce the theory, we have 
to make sure that this choice of gauge is maintained throughout the 
evolution of the system, i.e., we must require 
\begin{equation}\label{30}
\frac{\de }{\de t} (R- t^{1/3}) = \frac{\partial }{\partial t} (R- t^{1/3})
+ [R-t^{1/3} , H] = 0, 
\end{equation}
where $H$ is the unreduced Hamiltonian found in (\ref{3}), i.e., 
$H = - N (\frac{\pi_R^2}{24 R} -\frac{\pi_{\phi}^2}{2 R^3})
= - N(\frac{\pi_R^2}{24 t^{1/3}}- \frac{\pi_{\phi}^2}{2 t})$. This leads 
to $N = - 4 t^{-1/3}/\pi_R$. Solving the constraint for $\pi_R$, 
namely $\pi_R = - \sqrt{12}\pi_{\phi}/R$, we find $N = \frac{2}{\sqrt{3} 
\pi_{\phi}}$. Since $\pi_{\phi}$ is a constant of motion (see below) 
we  find 
that $N$ is constant, meaning that the specific time coordinate can be 
identified with what is usually called cosmological time. That is 
the reason for our specific gauge choice $R = t^{1/3}$, which, for the 
rest, does not fundamentally differ from the choice $R = t$ that has 
been used in \cite{isham}. 

We point out once again that the determination of $N$ is needed in 
order to determine the complete geometry. Leaving out the above step 
and directly passing over to the reduced theory would result in  a loss 
of information, since $N$ cannot be determined neither by the constraint, 
nor by the dynamical equations. And the knowledge of $N$ is certainly 
of physical interest. Without $N$ we cannot, for instance, determine the 
(four dimensional) curvature scalar. In electrodynamics, one can proceed 
similarly, namely, impose, e.g.,  the Coulomb gauge $A^{\mu}_{\ ,\mu} = 0$, 
and then require that the gauge is maintained in time, i.e., 
$[H,A^{\mu}_{\ ,{\mu}} ] = 0$. This leads to $\Delta A_0 = 0$. In contrast to 
the present case, however, this step is actually not required, because 
we do not need to know anything about $A_0$. As argued before, 
both the electric and the magnetic fields are fully determined by the 
constraints $\pi^{\mu}_{\ ,\mu} = 0$ and by the dynamical equations for the 
transverse fields. (The only 
purpose of the above step is to check for consistency, i.e., to check 
whether  one is actually allowed to impose $A^{\mu}_{\ ,{\mu}} = 0$.)

We can now reduce the Lagrangian (\ref{3}) by inserting $R = t^{1/3}$ and 
the solution of the constraint $\pi_R = - \sqrt{12} \pi_{\phi}/R$. We find 
\begin{equation}\label{31}
L = \pi_{\phi} \dot \phi -  \frac{2}{\sqrt{3} t} \pi_{\phi}, 
\end{equation}
or for the Hamiltonian 
\begin{equation}\label{32}
H =  \frac{2}{\sqrt{3} t} \pi_{\phi}.  
\end{equation}
Up to a factor $1/3$, this is the same Hamiltonian that one obtains in 
the gauge $R = t$, see \cite{isham}. From (\ref{31}), we see that 
the canonical variables are $\phi$ and $\pi_{\phi}$, and  consequently,  
we introduce the multiplication operator $\phi$ together with $\pi_{\phi} = 
- i \frac{\partial }{\partial \phi}$. Note by the way that (\ref{31}) 
leads to the classical solution $\pi_{\phi} = const$, and therefore, 
$N = const$, as outlined above. On the quantum level, the dynamics are 
described by the Schroedinger equation $H \psi(\phi,t) = i \frac{\partial}
{\partial t} \psi(\phi,t)$, i.e., 
\begin{equation}\label{33}
- i \frac{2}{\sqrt{3}t} \frac{\partial }{\partial \phi}\ \psi(\phi,t) = 
i \frac{\partial}{\partial t}\ \psi(\phi,t). 
\end{equation}
There is no factor ordering problem in that equation, and upon introducing 
a new variable $\tau = \sqrt{12} \ln t^{1/3}$, one finds 
$(\partial_{\phi} + \partial_{\tau}) \psi = 0 $, i.e., $\psi =
\psi(\phi-\tau)$. In other words, the general solution of (\ref{33}) 
reads 
\begin{equation}
\psi(\phi,t) = \psi(\phi- \sqrt{12} \ln t^{1/3}), 
\end{equation}
which, for $R = t^{1/3}$, is fully consistent with the results (\ref{18}) 
obtained from the direct reduction. In this  gauge, it is particularly 
obvious that the dynamical degree of freedom is contained in the scalar 
field and not in the geometry. 
Similar as in section \ref{reduc}, the reduction to (\ref{31}) is not 
unique (there are two roots to the constraint equation), and thus, 
the above solutions have to be completed by adding the contributions 
of the form $\psi(\phi,t) = \psi(\phi + \sqrt{12} \ln t^{1/3})$.

Since (\ref{33}) is free of ordering 
problems, we could have equally well used this particular gauge 
in order to determine the {\it correct} ordering  that has to be adopted 
in the Wheeler-DeWitt equation. The fact that in the specific 
gauge, the theory is free of ordering problems turns out to be 
a coincidence and is not a general feature, as we will see in our next 
example. 
  
Let us  consider the gauge $\phi = t$. Again, we have to require that 
the gauge is maintained throughout the evolution, similar as in (\ref{30}), 
before we solve the constraint. This is easily done and leads to 
$N = R^3/\pi_{\phi}$, which, for classical solutions, 
leads to an exponential behavior $N \sim \exp(\sqrt{3}t/2)$. Next, 
we reduce the Lagrangian, inserting $\phi = t$  and solving the 
constraint (conveniently for $\pi_{\phi}$ this time). The result is 
\begin{equation}
L = \pi_R \dot R - \frac{1}{\sqrt{12}}\ R \ \pi_R, 
\end{equation}  
and thus, the Hamiltonian reads 
\begin{equation} \label{36}
H = \frac{1}{\sqrt{12}}\ R \ \pi_R,
\end{equation} 
which is time independent this time, but not free of ordering 
problems. With the multiplication operator $R$ and $\pi_R = 
- i \partial_R$, we write the Schroedinger equation in the form 
\begin{equation}
- i \frac{1}{\sqrt{12}}\  R \frac{\partial }{\partial R}\ \psi(R,t) =  
i \frac{\partial}{\partial t}\ \psi(R,t), 
\end{equation} 
which has the general solution $\psi(R,t) = \psi(t - \sqrt{12} \ln R)$, 
consistently with (\ref{18}) and $t = \phi$. Note that this result 
has been obtained using the particular ordering (\ref{36}), and not, 
e.g., the symmetric ordering $(1/2)(R\  \pi_R + \pi_R R)$ that is 
conventionally adopted. Any ordering different from the above 
would lead to results that are not equivalent to 
those obtained in the  reduced theory or to those obtained in 
the gauge $R = t^{1/3}$, which were both free of ordering problems. 

One might argue that $H$ is not hermitian with the above ordering. 
This, however, is again related to the  question of the construction of 
the scalar product. Indeed, the above Hamiltonian suggests the use 
of the form 
\begin{equation}
<\psi_1 , \psi_2>\ = \int \psi_1^* (i \partial_R) \psi_2 \de R, 
\end{equation}
with suitable integration boundaries, at which the wave function is 
assumed to vanish. This product satisfies the usual properties of a 
scalar product except that it is not positive definite. We can 
now show that $H$ from (\ref{36}) satisfies the relation 
\begin{equation} 
<\psi_1, H \psi_2>\ = <H \psi_1, \psi_2>, 
\end{equation}
that is, $H$ is pseudo-hermitian. The same holds for $\pi_R$, but not 
for $R$. Nevertheless, $H$ has complex eigenvalues, as is easily shown 
by considering an eigenfunction $\psi_a$ satisfying $H\psi_a = 
a \psi_a$ and showing that the state $R \psi_R$ is again an eigenfunction 
of $H$, with the eigenvalue increased by an imaginary amount. This 
shows again the difficulties in the construction of the scalar product and 
in the interpretation of the wave function in general \cite{vilenkin}.

\section{Change of variables} \label{var}

In \cite{christo3}, the question has been raised whether one could 
eventually circumvent the difficulties with the ordering problems 
by multiplying the constraint with $R$ before the transition to the 
quantum theory. According to 
the authors, this is not allowed, because the resulting wave functions 
would exhibit radically different behaviors than the original ones. 
We will demonstrate that in fact, both parts of the 
statement are not  entirely accurate. Namely, we find that multiplication 
with $R$ does not resolve the ordering problems, and that the resulting 
wave equation has the exactly same solutions than in the original 
formulation, provided a consistent ordering is adopted. 

Multiplication of the constraint with $R$ is essentially the result 
of a change of variables. Namely, if we 
 introduce a new variable $\tilde N = \frac{N}{R}$, the  Lagrangian 
(\ref{3})  takes the form 
\begin{equation} \label{38} 
L = \pi_R \dot R + \pi_{\phi} \dot \phi + \tilde N 
( \frac{\pi_R^2}{24} - \frac{1}{2 R^2} \pi_{\phi}^2).  
\end{equation}
This can be explicitely reduced to 
\begin{equation} \label{39}
L =  \pi_{\phi}\dot X, 
\end{equation}
with $X$ as before, 
where we have solved the constraint assuming again that $\pi_R$ and 
$\pi_{\phi}$ have different signs (case 1). In other words, 
the dynamical variable is $X$, and the state function is a general 
function $\psi(X)$. Of course, just as in section \ref{reduc}, the 
solution of the constraint (and thus the reduction procedure) 
is not unique, and we have to add contributions  of the form 
$\psi = \psi(Y)$. 

Therefore, if the theory is still consistent, 
the Wheeler-DeWitt equation 
\begin{equation} \label{40}
\left(\frac{\pi_R^2}{24} - \frac{1}{2 R^2} \pi_{\phi}^2\right)\ \psi(R, \phi) 
= 0 
\end{equation}
should be identically satisfied for $\psi =  \psi(\phi - \sqrt{12} \ln R)$ 
(and for $\psi = \psi(\phi + \sqrt{12} \ln R)$ or a linear combination of 
both). 
At first sight, this does not seem possible, and it seems as if  
(\ref{40}) is free of ordering ambiguities. However, (\ref{40}) is 
indeed identically satisfied for the above solutions if we 
adopt the {\it ordering} $\pi_R^2 = \frac{1}{R} \pi_R R \pi_R$. 
This looks rather unorthodox, but it is  
actually very similar to the ordering in  (\ref{21}). 

We are thus led to the conclusion that $\pi_R^2$ is not 
an unambiguously defined operator. As a result, we must  
  assume the same for $\pi_{\phi}^2$, meaning that we actually have not 
been very careful in section \ref{wdw}, where we have simply assumed 
that $\pi_{\phi}^2 = \pi_{\phi} \pi_{\phi}$. It is, however, obvious 
that this is indeed the only possible ordering in the second 
term of (\ref{21}) in order for  (\ref{20}) to be identically satisfied. 

A more clever choice than just replacing $N$ is by using the 
variables $X$, $Y$ and $\tilde N = 2 R^{-3} N$. 
The second order Lagrangian (\ref{1}) then 
simplifies to $L = \frac{1}{\tilde N} \dot X \dot Y$, and the 
corresponding first order Lagrangian reads 
\begin{equation} \label{41}
L = \pi_X \dot X + \pi_Y \dot Y + \tilde N \pi_X \pi_Y. 
\end{equation}
The analysis is now trivial. We solve the constraint 
by $\pi_Y = 0$ (case 1) or $\pi_X = 0$ (case 2), and 
Lagrangian reduces to (in case 1) $L = \pi_X \dot X$, with dynamical 
variable $X$ and wave functions $\psi(X)$. On the other hand, in 
the Wheeler-DeWitt approach, we have to solve the equation 
\begin{equation}\label{42}
(\pi_X \pi_Y)\  \psi(X,Y) = 0,  
\end{equation}
which is, this time, completely free of ordering problems, and 
leads directly to the solutions $\psi = \psi_1(X) + \psi_2(Y)$, 
just as in section \ref{wdw}. 

The above choice of variables is essentially the only one that 
leads to a Wheeler-DeWitt equation without ordering problems (without 
explicitely fixing the gauge).  
It is thus clear that in general, where we do not know the 
explicit solutions of the field equations, 
we will not be able to find the corresponding form of the 
Wheeler-DeWitt equation, meaning that such a change of
variables cannot be used as a tool 
to determine the correct ordering that has to be adopted when 
the same equation is formulated in terms of other variables. The argument
nevertheless demonstrates once more  that there is indeed a unique 
factor ordering in order to get a consistent theory, 
and that it is not simply  a matter of a global shift in the 
 energy level shifts, as has been suggested, e.g.,  in \cite{isham}.   

On the occasion, we also observe that it is not possible to 
fix the gauge implicitly by imposing a condition 
$f(X,Y,\pi_X, \pi_Y)= 0$ with a time independent $f$. Indeed, stabilization 
of such a gauge choice will always lead to an expression proportional 
to $\tilde N$, which would mean that either $\tilde N$ has to be 
zero (which we have to exclude), or that we get yet another 
condition $g(X,Y,\pi_X,\pi_Y) = 0$ on the variables. Together with 
the original Wheeler-DeWitt constraint, that  would make three relations 
between four variables, which is obviously not allowed, since 
two variables correspond to 
 propagating  fields. Thus, any gauge fixation will necessarily 
depend explicitely on the time coordinate, $f(X,Y,\pi_X,\pi_Y,t) = 0$. 
Since there is no reason to involve the dynamical fields into 
such a condition, and since 
$\pi_Y$ is already zero on the constraint (in case 1), 
the most natural choices 
are of the form $Y = f(t)$ for some function $f(t)$. Stabilization leads 
to $\tilde N^{-1} = - \pi_X/\dot f(t)$. 
Conditions of the form $Y + g(X) = f(t)$ lead to 
the same expression for $\tilde N$, since the Poisson 
bracket of the second term 
with the Hamiltonian is proportional to $\pi_Y$, which 
is zero on the constraint surface. In fact, it is obvious that 
the dynamical variables commute with the constraint anyway, and there is 
thus no need to include them into the gauge fixing conditions. 
Nevertheless, in this way one can formally obtain, e.g., 
the previously used gauges $\phi = t$ or $R = t^{1/3}$. 
It should be noted that the above argumentation assumes that 
the solutions are of type one. In the case where $Y$ is the 
dynamical field, we can obviously not use $Y = f(t) $, but 
should rather use $X = f(t)$. One might therefore think that a better
choice would be of the form, e.g.,  $X + Y = f(t)$, since it 
covers both cases. This does not solve all the problems, though, as we 
will see in the next section. 

The fact that the operator ordering can be fixed by performing a 
change of variables such that the resulting equations take a form 
free of ordering problems was also the starting point of the 
investigations carried out in \cite{christo2,christo3}. In particular, 
the authors managed to generalize their arguments to the complete theory.

The form (\ref{42}) is also particular convenient to analyze the 
classical limit of the theory. Namely, if we write the 
wave function in the form $\psi = a(X,Y) \exp[i S(X,Y)/\hbar]$, with 
real functions $a$ and $S$, then the Wheeler-DeWitt equation leads to 
\begin{eqnarray}\label{59}
\frac{\partial S}{\partial X} \frac{\partial S}{\partial Y}  &=& 
\frac{\hbar^2}{a} \frac{\partial^2 a}{\partial X \partial Y} \\ \label{60}
\frac{\partial S}{\partial Y} \frac{\partial a }{\partial X}
+\frac{\partial S}{\partial X} \frac{\partial a }{\partial Y}
&=& - a \frac{\partial^2 S}{\partial X \partial Y} . 
\end{eqnarray}
The first equation, in the limit $\hbar \rightarrow 0$, reduces to the 
Hamilton-Jacobi equation, while the second equation is commonly interpreted 
as conservation equation for the probability density. We see in particular 
that  the pure states of the 
form $\psi = \psi(X)$ or $\psi = \psi(Y)$ are not 
only solutions of (\ref{59}) and (\ref{60}), but they are also 
exact solutions of the Hamilton-Jacobi equation 
$\frac{\partial S}{\partial X} \frac{\partial S}{\partial Y}  =0$.  
Only for mixed solutions $\psi= \psi(X) + \psi(Y)$ will the 
quantum corrections in (\ref{59}) be effective.  
The same analysis is easily performed in the original variables 
$R$ and $\phi$, where one can also directly verify  that 
the Hamilton-Jacobi limit does not depend on the operator ordering.

\section{Discussion}

The analysis of the simple cosmological models allows us to 
identify a direct consequence of the fact that 
 the theory is not completely determined by 
the constraints and by the dynamical equations alone. That this is 
the case, we repeat, can simply be recognized by the fact that 
we have 5 variables in the theory, $\phi,\pi_{\phi},R,\pi_R,N$, 
two of which are dynamical, one is eliminated by the constraint, 
and one can be fixed by imposing a gauge. This leaves one undetermined 
variable. However, once the time coordinate has been chosen, the 
metric should  be determined completely. The {\it missing} relation can 
only be obtained from the unreduced Hamiltonian, for instance by 
stabilization of the gauge fixing condition. 
But this turns out to 
lead to an astonishing feature of the theory. 

Indeed, since we concentrated on the dynamical variables, we have glanced 
over a particularity  of the theory. 
To be explicit, 
we consider the case where the time coordinate is chosen  as 
$R= t^{1/3}$, but similar arguments apply to any other gauge choice.  

We have shown that  stabilization of the gauge $R= t^{1/3}$ leads to 
$N = \frac{2}{\sqrt{3} \pi_{\phi}}$, from which we concluded that, 
since $\dot \pi_{\phi} = 0$, $N$ has to be a constant. This is true 
on the classical level, but it is not entirely 
accurate in the quantum theory. Indeed, since in the specific gauge, 
$\phi$ and $\pi_{\phi}$ are the dynamical fields of the theory, we have 
to interpret the above relation as an operator equation. Little 
attention has been given to this particularity of gravity in the literature, 
although it has been explicitely mentioned in \cite{dewitt}. Multiplying 
from the right with $\pi_{\phi}$ and from the left with $N^{-1}$, we can 
write 
\begin{equation}\label{61}
N^{-1} = \frac{\sqrt{3}}{2} \pi_{\phi} = - i \frac{\sqrt{3}}{2} 
\frac{\partial }{\partial \phi}. 
\end{equation}
Similarly, in the variables of the previous section, and the gauge 
$Y = f(t)$, one obtains the operator $\tilde N^{-1} = i \dot f^{-1} 
\frac{\partial}{\partial x}$. 
Thus, although, for instance for the above choice, 
 $\pi_{\phi}$ (and thus $N$) is  constant in the classical theory, 
and we also have  
$i [H,\pi_{\phi}] =  \dot \pi_{\phi}  = 0  $ in the Heisenberg representation 
(with $H$ from (\ref{32})), the  statement that $N$ is constant ({\it
cosmological time}) does  not make sense in the Schroedinger picture. 

The fact that there is after all a component of the metric  that does not 
behave classically should not be surprising. Since in the specific gauge, 
the only dynamical variables are $\phi$ and $\pi_{\phi}$, while $R$ is 
fixed by the gauge choice, it would be rather strange if the remaining 
variable contained in the metric (\ref{2}) would be classical also. 
A back-reaction of spacetime to the presence of the quantum field 
$\phi$ should naturally be expected, and it turns out, in our specific case, 
that this is achieved not by dynamical (quantum) degrees of freedom 
of the metric (i.e., by the presence of {\it gravitons}), but rather 
by the quantum nature of the lapse  function. 

It should be noted that this feature is very unique: We have a quantum 
operator $N$ that has to be determined by purely classical means, 
namely by variation of the Lagrangian (prior to solving the constraint), 
or equivalently by 
equation (\ref{30}) (recall that (\ref{30}) is a classical equation, 
and $[H,R]$ is the Poisson bracket). For instance, there is no Heisenberg 
equation $i[H,N] =  \dot N $ prior to the determination of $N$ in terms 
of $\pi_{\phi}$. One can argue that the same holds, e.g., for $\pi_R$, 
which is also determined classically by solving the constraint, but 
this is a different situation. $N$ is not a constrained variable, 
neither a dynamical one (in the strict sense) and it 
is nevertheless needed to determine the configuration completely.  
As to $\pi_R$,  it can either be eliminated prior to quantization, 
or, in the Wheeler-DeWitt approach, it remains in the theory, but 
its relation to the other variables is not given in terms of 
an operator equation (i.e., the constraint $\frac{\pi_R^2}{24R} - 
\frac{\pi_{\phi}^2}{2 R^3} = 0$ applies only to physical states). On 
the other hand, equation (\ref{61}) is a true operator equation 
(see the remarks in \cite{dewitt} on this point).

This particular situation is obviously a result of the fact that 
the theory is not described merely by the dynamical variables and the 
constraints and cannot occur in conventional gauge theories. For instance, 
in electrodynamics, it is possible that  $A_0$ turns into a quantum 
operator, for instance through the stabilization of the gauge fixing 
$A^{\mu}_{\ ,{\mu}} =0$, which leads in the presence of 
charged matter to $\Delta A_0 = \rho$. Since $\rho$ 
is gauge invariant, it has to be constructed from  
 the dynamical variables of the matter field, and thus from operators in 
the quantum theory, e.g.,  $\rho = e 
\psi^{\dagger}\psi$. However, this is not  
physically relevant, since the knowledge of $A_0$ is not 
needed for the description of the theory (what we need instead is 
the constraint $\pi^{\mu}_{\ ,\mu} - \rho = 0$, which is, however, not 
an operator relation and applies only to the physical states) 
and thus we can simply forget about $A_0$.  

Note that the operator form of $N$ can only be determined once a 
specific gauge has been chosen. Prior to this, there is no expression 
of $N$ in terms of the remaining fields that does not involve velocities.
(Obviously, since if there were, 
this would mean that there is an additional constraint 
in the theory.) What we have (for instance in case 1), is an 
equation of the form $\dot Y = \frac{2N}{R^3} \pi_{\phi}$ (on the 
constraint surface). Only if we fix the gauge variable $Y$ 
such that $\dot Y$ becomes an explicit  function of the 
field  variables and of $t$ 
can we derive the explicit form of the operator $N$. 

We should point out that there is actually an additional problem. 
In fact, since the reduction is not unique, and we actually cannot 
decide which are the physical variables, we do not really know 
which are the allowed gauge choices.  As pointed out before, 
the choice $Y = f(t)$ is only possible in case 1. But even if 
we assume, for instance, that the choice $R = t^{1/3}$ that we 
used above, is allowed in both cases, leading to $\phi$ and $\pi_{\phi}$ 
as physical variables, then we still cannot determine the operator 
form of $N$ without explicit reference to a specific type of solutions. 
Namely, fixation of the gauge condition leads, as we have seen, to 
$N = - 4 t^{-1/3}/ \pi_R$, where $\pi_R$ has to be eliminated with the 
help of the constraint. This cannot be done uniquely, and leads to 
both solutions $N^{-1} = \pm i \frac{\sqrt{3}}{2} 
\frac{\partial }{\partial \phi}$, compare with (\ref{61}). In this 
particular case, we can simply claim that the ambiguity in the sign is 
not of any relevance, since the original variable in the metric was 
$N^2$ anyway. But if we consider, for instance, the variables used 
in the previous section, and fix the gauge by $X + Y = f(t)$ 
(which is allowed, since it does  not constrain the dynamical  
variable neither in case 1 nor in case 2), then stabilization leads to 
$\tilde N^{-1} = - (\pi_X + \pi_Y)/ \dot f$. One of both momenta 
is to be eliminated by the constraint, but we cannot determine which one,  
meaning  that we do not know the explicit operator form of 
$\tilde N^{-1}$ before we identify either $X$ or $Y$ as physical variable 
(i.e., before we confine ourselves to a particular type of classical 
configurations). 

In particular, although the Wheeler-DeWitt approach was previously very 
elegant, since it allowed us to find directly the general form of the 
wave functionals 
without solving the constraint classically, and thus without choosing 
one of both roots, the problem reappears at the level of the determination 
of the operator form of $N$. As pointed out before, $N$ can only be 
determined after a gauge has been chosen, since prior to this, 
there is not equation for $N$ that does not involve time derivatives. 
But if we fix  the gauge, we are left with three variables, and thus, 
we cannot work with two pairs of canonically conjugated variables 
(as in the Wheeler-DeWitt approach).
Rather, we have to 
 eliminate one of the remaining variables with the help of the constraint and 
reduce the theory to a single pair of canonically conjugated variables. 

Alternatively, one could  
argue that we do not need the operator form of 
$N$. Since we know already the solutions for the wave function, 
one could claim that anything else  should be determined with the help 
of that wave function. Such an argumentation, however, is hard to support, 
since we already know that, on the classical level, we need to know 
$N(t)$ (once the gauge is fixed), and therefore, if we omit it completely 
on the quantum level, then we also loose the possibility to 
recover the classical limit. In fact, in \cite{gerlach}, a theorem has 
been proven  which states that from the Hamilton-Jacobi limit of the 
Wheeler-DeWitt equation, we can derive the classical Hamiltonian equations 
for the canonical fields $\pi^{\mu\nu}$ and $g_{\mu\nu}$, and that, if 
those are satisfied, there exist functions (lapse and shift) 
$N$ and $N_{\mu}$ such that the metric $g_{ik}$ (constructed in the usual 
way from $g_{\mu\nu}$ and $N,N_{\mu}$) satisfies the Einstein field
equations. The functions $N$ and $N_{\mu}$ themselves, however, are not 
specified by the Hamilton-Jacobi equation. (Meaning that, even if we 
fix the coordinate system by imposing conditions on the non-propagating 
components of $\pi^{\mu\nu}$ and $g_{\mu\nu}$, the metric will not be 
determined completely.)

Finally, we present a few speculations concerning  the reason 
for those specific 
features of  General Relativity. One could  think that 
this feature of gravity is a result of the reparameterization invariance. 
However, as we have seen,  we have a similar situation in 
the special relativistic spin-two theory, which is not reparameterization  
invariant, and on the other hand, the special relativistic 
point-particle 
Lagrangian, which is reparameterization invariant, does not share the 
same features, see \cite{leclerc}. 
In fact, in the latter  case, the role of $N$ is played by 
a Lagrange multiplier $\lambda$, which is introduced by hand and is thus  
trivially unphysical. Thus, if we chose, e.g., the gauge $\tau = t$ 
($\tau $ is the parameter of the theory, $S = \int L \de \tau$, see 
\cite{leclerc}), then the stabilization  of the constraint leads merely to the 
determination of $\lambda$, which is completely uninteresting, while 
the dynamical variables determine the position $x^{\mu}$ in terms of $t$. 

Since the reason cannot be found neither in the general covariance, 
nor in the reparameterization invariance, nor in the non-linearity 
of the theory, where can it be found? The only common feature of 
the linear spin-two theory and General Relativity we can 
think of is the fact that they are spin-two theories. By (massless) spin-two 
theory, we mean a theory whose first order approximation leads to 
a wave equation for a symmetric, traceless-transverse tensor field, 
and is, in addition, Lorentz invariant. Note that in the case of 
General Relativity, this requires reference to a fixed  background metric. 
We believe that the reason for the fact that both General 
Relativity as well as the linear theory are not completely described 
by the constraints and the dynamical variables  (i.e., by the spin-two field) 
alone is the fact that they are {\it more than just spin-two theories}, 
that is, they are not determined by the mere fact that they are 
spin-two theories. Otherwise stated, the spin-two theory is not unique. 
This is obvious, since we have already mentioned two different theories, 
but there are actually more. For instance, one could start from a 
special relativistic theory based on a traceless field $h_{ik}$ 
(traceless in four dimensions, in order to retain Lorentz invariance). 
Similarly, a modification of General Relativity based on 
a metric that satisfies  $\det g_{ik} = -1$ 
(the so-called unimodular theory, see \cite{henn,unruh})
still leads to a spin-two theory. 

The important thing is that those 
theories are all based on different gauge groups, meaning that the 
requirement of a theory to describe a spin-two particle does not 
fix uniquely the symmetry group. On the other hand, there is only 
one possible gauge group for the spin-one theory. Indeed, if 
we stick to linear theories, then the only possible form of the 
reduced Hamiltonian is given by (\ref{m3}), and the only Lorentz 
invariant theory that leads to (\ref{m3}) is the conventional Maxwell 
theory. (Of course, one could start with a field satisfying the 
Lorentz gauge $A^i_{\ ,i} =0 $ from the outset, but that leads to 
an equivalent theory.) Moreover, it has been shown in \cite{wald} 
that the only non-linear generalizations  of Maxwell theory necessarily 
admit the same gauge group as the linear theory. (Only if we admit 
several spin-one fields, as is the case in Yang-Mills theory, will the gauge 
group be modified by higher order terms.)  

In the spin-two case, the situation is different. Even if 
we start from the symmetry group of the conventional theory, 
namely $\delta h_{ik} = \xi_{i,k} + \xi_{k,i}$, the non-linear 
generalizations can admit either the same symmetry transformations, 
or alternatively the symmetry transformations of a generally covariant 
theory, see {\cite{wald}. Moreover, one can start from a linear theory 
with a reduced symmetry, like the one mentioned before (where $h_{ik}$ 
is traceless and $\xi_i$ has to satisfy $\xi^i_{\ ,i} = 0$) and still 
end up with a spin-two theory, which is, however, not equivalent to 
the original one. Thus, although in the conventional theory, 
we can always choose a (Lorentz covariant) gauge where $h^i_{\ i} = 0$, 
imposing this from the outset does not lead to the same theory, 
quite in contrast to the above mentioned gauge $A^i_{\ ,i} = 0$ in 
electrodynamics. The non-linear extension of such a traceless theory 
leads to theories admitting either the same gauge group, or the 
gauge group of unimodular gravity. Note that in General Relativity, 
the situation is similar to the linear theory, 
namely we can always choose a coordinate 
system such that $\det g = -1 $, but imposing this from the start 
(i.e., working with a reduced number of fields) leads to a 
different, inequivalent theory. (We should note, however, that in the 
linear case, the theory based on the traceless field, although 
mathematically different, is physically 
equivalent to the conventional theory, since the reduced Bianchi identities 
fix  the trace  of the field equations to a constant. With the conventional 
boundary conditions of special relativistic theories, this constant can only 
be zero anyway. In the non-linear case, the situation is different and we 
obtain, in the unimodular case,  an unspecified cosmological constant.) 

We believe that this non-uniqueness of the spin-two theory is 
the reason for which the theories are not  determined by the constraints 
and by the dynamical equations alone. It is not clear to us though 
whether this is due to the fact that already a single linear spin-two 
theory admits non-linear generalizations with two different symmetry  groups 
(which is the  non-trivial  result of the analysis 
carried out in  \cite{wald}), or whether it is due to the much 
simpler fact that there are already at the linear level several 
spin-two theories.

\section{Conclusions}

We have shown that the special relativistic spin-two theory is not 
completely specified by the constraints and by the equations for the 
dynamical fields alone. We argued that the same holds in General Relativity, 
and have shown this for simple cosmological examples for which  the 
Hamiltonian reduction can be performed explicitely. While this fact 
alone is not surprising and is a general feature of gauge theories, 
in the above cases, we have the situation that there exist gauge invariant 
expressions formed with the help of the unspecified variables that cannot 
be determined without reference to the initial, unreduced Hamiltonian. 
Thus, in contrast to, e.g., electrodynamics, where both the electric and 
the magnetic fields are specified by the constraint and by the gauge invariant 
(transverse) dynamical fields, in the above theories, there are additional 
gauge invariant expressions that are not determined by the constraints  
and contain nevertheless the non-dynamical variables. This concerns, in 
the spin-two theory, the trace of the Lagrangian field equations (expressed 
in terms of Hamiltonian variables), and in General Relativity the 
four dimensional curvature scalar ${}^{4}\!R$ (expressed in terms of   
 Hamiltonian variables). In other words, the invariant 
equation ${}^{4}\!R = 0$ cannot be expressed in terms of 
the dynamical fields alone (it contains, e.g., the lapse function $N$), 
nor is it a constraint (it contains, e.g., the velocity $\dot \pi$). 
It can therefore not be obtained from the explicitly reduced Lagrangian, 
as we have shown on the example of Robertson-Walker cosmologies. The situation 
is not changed if we fix the coordinate system. That is, if we impose 
gauge conditions on the non-dynamical components of $g_{\mu\nu}$ and 
$\pi^{\mu\nu}$, the functions $N_{\mu}$ and $N$ that appear in the expression 
for ${}^4\!R$ still remain undetermined. The only way to determine those 
functions (and thus to obtain the complete set of field equations) is to 
refer to the initial, unreduced Hamiltonian. Therefore,  the equations 
that determine those  functions 
have to be derived {\bf before} we actually identify the physical
(propagating) variables of the theory. If we stick to the conventional 
procedure that only propagating variables have to be quantized this means 
that the above functions have to be determined by their {\it classical}
field equations. Once we have established the form  
$N$ and $N_{\mu}$  in terms   of the remaining fields, 
we can proceed with  the Hamiltonian  reduction,  
 identify the propagating fields and quantize the theory. However, since 
$N$ and $N_{\mu}$ are now given as expressions in terms of propagating fields
(which we can only interpret as operator equations), we have the 
strange situation that those components are after all quantum operators. 
In summary, we have  {\it quantum} operators whose explicit form can only be 
determined by solving {\it classical} equations of motion. While all the 
above concerning $N$ and $N_{\mu}$ holds in exactly the same way for, 
e.g., $A_0$ in electrodynamics, the important thing 
is that we cannot simply discard the variables $N$ and $N_{\mu}$ as 
unphysical gauge variables (as we can  with $A_0$), 
since they are needed in the classical theory to 
obtain the complete set of coordinate invariant quantities (e.g., the scalar 
${}^4\!R$). If we require the quantum theory to reduce to 
the classical one in an appropriate limit, it is clear that we need the 
operator form of $N$ and $N_{\mu}$. To be precise, we have shown that 
the above arguments hold for $N$. It is possible that the expression for 
$N_{\mu}$ is not needed, i.e., that for $N_{\mu}$ we have a situation similar 
to conventional gauge theories, meaning that there is no gauge invariant 
relation that contains $N_{\mu}$ other than those that are already 
determined by the constraints and the propagating equations. 
(As mentioned before, in electrodynamics, the only invariants are 
the electric and the magnetic fields, and those are determined completely 
by the constraint $\pi^\mu_{\ ,\mu} = 0$ and by the equations for 
the transverse fields $\P^{\mu}$ and $\AA_{\mu}$. No equation containing 
$A_0$ is thus needed.) 

Further, we have shown  for the simple cosmological models considered 
in our analysis that in order to obtain consistency between the 
explicitely reduced theory and the Wheeler-DeWitt approach, a very 
specific operator ordering has to be adopted in the Wheeler-DeWitt equation. 
In contrast to previous arguments, we obtained 
our  result  without reference to hermiticity properties (and thus to 
the scalar product) and 
without reference to  the classical limit. Although this line or argumentation 
cannot  directly be generalized to the full theory (since the 
explicit reduction is not possible), it shows nevertheless that the discussion 
could eventually be carried out independently of the concrete construction 
of the scalar product.

\end{document}